\documentclass[reprint,showpacs,preprintnumbers,amsmath,amssymb,longbibliography,aip]{revtex4-1}
\usepackage[utf8]{inputenc}
\usepackage{mathrsfs}
\usepackage{amsmath}
\usepackage{bm}
\usepackage{amsfonts}
\usepackage{amssymb}
\usepackage{amsthm}
\usepackage{color}
\usepackage{graphicx}
\graphicspath{{./images/}}
\usepackage{geometry}
\usepackage{hyperref}
\usepackage{ulem}
\usepackage{epstopdf}
\usepackage{graphicx}
\hypersetup{
     unicode=false,
     pdftoolbar=true,
     pdfmenubar=true,
     pdffitwindow=false,     % window fit to page when opened
     pdfstartview={FitH},    % fits the width of the page to the window
     pdftitle={My title},    % title
     pdfauthor={Author},     % author
     pdfsubject={Subject},   % subject of the document
     pdfcreator={Creator},   % creator of the document
     pdfproducer={Producer}, % producer of the document
     pdfkeywords={keyword1} {key2} {key3}, % list of keywords
     pdfnewwindow=true,      % links in new PDF window
     colorlinks=true,       % false: boxed links; true: colored links
     linkcolor=blue,          % color of internal links (change box color with linkbordercolor)
     citecolor=blue,        % color of links to bibliography
     filecolor=magenta,      % color of file links
     urlcolor=blue           % color of external links
}
\usepackage[toc,page]{appendix}
\begin{document}
\title{Current-driven motion of magnetic domain-wall skyrmions}
\author{Haoyang Nie}
\affiliation{School of Physics and Electronics, Hunan University, Changsha 410082, China}
\author{Zhixiong Li}
\affiliation{School of Physics, Central South University, Changsha 410083, China}
\author{Xiansi Wang}
\affiliation{School of Physics and Electronics, Hunan University, Changsha 410082, China}
\author{Zhenyu Wang}
\email[Corresponding author: ]{vcwang@hnu.edu.cn}
\affiliation{School of Physics and Electronics, Hunan University, Changsha 410082, China}

%\date{\today}% It is always \today, today, but any date may be explicitly specified

\begin{abstract}
Domain-wall skyrmions (DWSKs) are topological spin textures confined within domain walls that have recently attracted significant attention due to their potential applications in racetrack memory technologies. In this study, we theoretically investigated the motion of DWSKs driven by spin-polarized currents in ferromagnetic strips. Our findings reveal that the motion of DWSKs is contingent upon the direction of the current. When the current is applied parallel to the domain wall, both spin-transfer torque (STT) and spin-orbit torque (SOT) can drive the DWSK along the domain wall. Conversely, for currents applied perpendicular to the domain wall, STT can induce DWSK motion by leveraging the skyrmion Hall effect as a driving force, whereas SOT-driven DWSKs halt their motion after sliding along the domain wall. Furthermore, we demonstrated the current-driven motion of DWSKs along curved domain walls and proposed a racetrack memory architecture utilizing DWSKs. These findings advance the understanding of DWSK dynamics and provide insights for the design of spintronic devices based on DWSKs.
\end{abstract}

\maketitle

Magnetic skyrmions are topological spin structures in chiral magnets and have emerged as a subject of immense interest for their potential as information carriers in spintronic devices \cite{Nagaosa2013,Fert2017,Sitte2018,Zhou2019,Kang2019,Back2020,Zhang2020,Du2022,Zhang2023}.
In particular, the concept of skyrmion-based racetrack memory has attracted considerable attention, where the data bits ``1" and ``0" are respectively encoded by the presence and absence of a skyrmion \cite{Kang2019,Sampaio2013,Fert2013,Tomasello2014,Kang2016}.
However, the application of magnetic skyrmions faces certain limitations.
Firstly, the skyrmion motion is affected by the skyrmion Hall effect \cite{Jiang2017,Litzius2017,Chen2017}, which causes the skyrmion trajectory to deviate from the racetrack direction due to the Magnus force. At high velocities, this effect can lead skyrmions to move toward the racetrack edges, resulting in their annihilation and loss of information.
Secondly, the data representation in skyrmion-based racetrack memory is susceptible to various factors, such as thermal fluctuations \cite{Barker2016,Zazvorka2019,Zhao2020}, which cause variable spacing between skyrmions. Consequently, the accurate encoding of the data bit ``0" becomes challenging.

To overcome the aforementioned limitations, a range of strategies have been proposed. For instance, the skyrmion Hall effect can be mitigated by adding additional energy barrier at the nanotrack edges \cite{Lai2017}, employing a strip domain wall as a buffer layer \cite{Xing2020}, modifying the effective spin torque \cite{Gobel2019}, and replacing the skyrmions by other topologically compensated hybrid skyrmions that possess zero topological charge, such as bilayer skyrmions \cite{Zhang2016}, antiferromagnetic skyrmions \cite{Zhang201602}, and skyrmioniums \cite{Gobel201902}.
On the other hand, the data representation challenge in skyrmion-based racetrack memories can be addressed through the use of a complementary racetrack with voltage manipulation \cite{Kang201602}, a two-lane racetrack \cite{Muller2017}, and a skyrmion-skyrmionium racetrack \cite{Li2023}.
Nevertheless, these approaches often entail increased design and fabrication complexity, which impede their practical application. Therefore, it is imperative to explore alternative information carriers for racetrack memories.

Recently, a composite object known as the domain-wall skyrmion (DWSK) has been studied \cite{Cheng2019,Ross2023,Amari2024,Han2024,Gudnason2024} and observed in chiral magnets \cite{Li2021,Yang2021}.
It is topologically equivalent to the conventional skyrmion and can be viewed as a skyrmion trapped inside a domain wall. Although the DWSK also displays the skyrmion Hall effect due to its nonzero topological charge \cite{Han2024}, its movement is restricted to the domain wall, effectively mitigating the skyrmion Hall effect.
Unlike conventional skyrmions, which can only exist in a ferromagnetic phase with a single topological charge, the DWSKs with opposite topological charges ($Q=\pm1$) can coexist within one domain wall. Thus, the data bits ``1" and ``0" can be respectively encoded as the DWSKs with $Q=\pm1$, providing an excellent solution to the data representation problem in conventional skyrmion-based racetrack memories.

Manipulating a DWSK is crucial for its applications. Spin-polarized currents are widely used stimuli to drive magnetic textures \cite{Beach2006,Caputo2007,Shen2019,Zeng2020,Hu2021,Wang2019,Ohki2025}.
In this study, we explore the dynamics of DWSKs driven by the spin-polarized current through spin-transfer torque (STT) \cite{Zhang2004} and spin-orbit torque (SOT) \cite{Liu2011}. Owing to the DWSK's nontrivial topology, its motion exhibits the skyrmion Hall effect, which can be counteracted by the confinement of the domain wall and the repulsion from the strip boundary, resulting in a stable motion along the domain wall.
We also demonstrate that DWSKs can move within a curved domain wall. The domain wall can serve as a channel to guide the motion of DWSKs and suppress the skyrmion Hall effect. Therefore, the DWSKs are superior to skyrmions as information carriers in racetrack memories.

\begin{figure}
  \centering
  \includegraphics[width=0.5\textwidth]{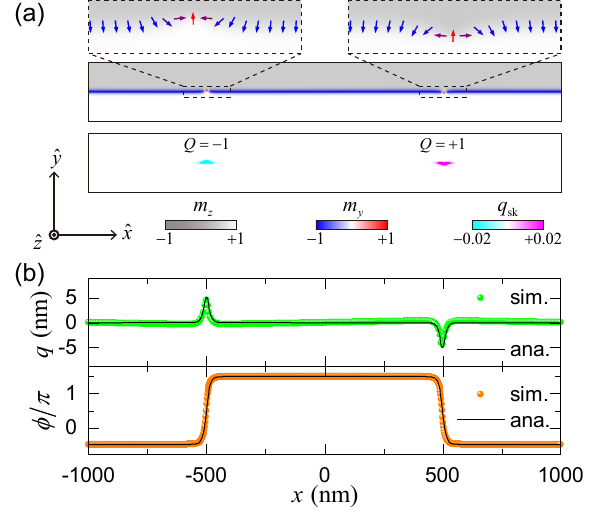}\\
  \caption{(a) The magnetization state of the DWSKs with opposite topological charges. $q_{\mathrm{sk}}$ is the topological charge density. (b) The position of the domain wall and the azimuthal angle of the magnetization inside it. The dots are simulation data and the curves are analytical results.}\label{fig1}
\end{figure}

We consider a ferromagnetic strip with the free energy given by
\begin{equation}\label{eq_energy}
  E=L_z \int\bigg\{A(\nabla\mathbf{m})^2-D\mathbf{m}\cdot\Big[(\hat{z}\times\nabla)\times\mathbf{m}\Big]-Km_z^2 \bigg\}dxdy
\end{equation}
where $\mathbf{m}=\mathbf{M}/M_s$ is the normalized magnetization with the saturated magnetization $M_s$, $L_z$ is the film thickness, $A$ is the exchange constant, $D$ is the coefficient of the interfacial Dzyaloshinskii-Moriya interaction (DMI), $K=K_u-\mu_0 M_s^2/2$ is the effective perpendicular magnetic anisotropy with $K_u$ the uniaxial anisotropy along the $z$ direction.
A N\'{e}el-type domain wall containing two DWSKs is located in the center of ferromagnetic strip, as depicted in Fig. \ref{fig1}(a).
For simplicity, we disregard the local demagnetization energy around the domain wall, which is accounted for in Ref. \onlinecite{Cheng2019} to ascertain the critical DMI strength ($D_c=2\ln2 L_z \mu_0 M_{s}^2/\pi^2$) necessary for the formation of a complete DWSK. In this study, we consider a large DMI constant ($D>D_c$) to support the DWSK, rendering the local demagnetization energy around the domain wall negligible---a fact that is later substantiated by micromagnetic simulations.

To describe the spin configuration of the DWSK, we express the unit magnetization in spherical coordinates as $\mathbf{m}=\{\sin\theta\cos\phi,\sin\theta\sin\phi,\cos\theta\}$, and introduce $q(x)$ and $\phi(x)$ to denote the cusp formed at the position of the DWSK and the azimuthal angle of domain wall magnetization, respectively. To be more specific, we set $D>0$ and $\partial_y\theta>0$, thus the magnetization within the domain wall is along $-\hat{y}$ direction. Then, the profile of the DWSK can be analytically solved by minimizing the total free energy, which is given by
\begin{subequations}\label{eq_profile}
\begin{align}
&\theta(x,y)=2\arctan\Bigg[\exp\bigg(\frac{y-q(x)}{\lambda}\bigg)\Bigg],\\
&\phi(x)=4\arctan\Bigg[\exp\bigg(\pm\frac{x}{\lambda_s}\bigg)\Bigg]-\frac{\pi}{2},\\
&q(x)=\lambda^2\partial_{x}\phi=\pm2\kappa\lambda_s\operatorname{sech}\bigg(\frac{x}{\lambda_s}\bigg),
\end{align}
\end{subequations}
where $\lambda=\sqrt{A/K}$ is the domain wall width, and $\lambda_s=\lambda/\sqrt{\kappa}$ is the size of DWSK with $\kappa=\pi D/(4\sqrt{AK})$.
Then, the topological charge of the DWSK can be calculated as
\begin{equation}\label{eq_Q}
  Q=\frac{1}{4\pi}\iint\Big(-\sin\theta\partial_{y}\theta\partial_{x}\phi\Big)dxdy=\mp1.
\end{equation}
The sign of ``$\pm$" in Eq. (\ref{eq_profile}b) represent clockwise and counterclockwise rotation of the magnetization in the DWSK, respectively.
From Eqs. (\ref{eq_profile}c) and (\ref{eq_Q}), one can see that both the cusp ($q$) and topological charge ($Q$) depend on the magnetization rotation ($\partial_{x}\phi$). At the position of the DWSK, the cusp points upward for $\partial_{x}\phi>0$ and downward for $\partial_{x}\phi<0$, corresponding to $Q=-1$ and $Q=1$, respectively.

To confirm the above theoretical analysis, we perform micromagnetic simulations using MuMax3 \cite{Vansteenkiste2014}. A ferromagnetic strip with the length of $L_x=2000$ nm, width of $L_y=250$ nm, and thickness of $L_z=2$ nm is considered. Magnetic parameters of Co are used in simulations: $A_{ex}=15$ $\mathrm{pJ/m}$, $M_s=5.8\times10^5$ $\mathrm{A/m}$, $D=1$ $\mathrm{mJ/m^2}$, $K_{u}=8\times10^5$ $\mathrm{J/m^3}$, and $\alpha=0.1$. The cell size is $2\times2\times2$ $\mathrm{nm^3}$. Periodic boundary conditions are used in the $x$ direction. Figure \ref{fig1}(a) shows the magnetization profile of the DWSK, obtained from simulations. One can see that DWSKs with opposite topological charges ($Q=\pm1$) coexist in one domain wall. By fitting the polar angle of magnetization using Eq. (\ref{eq_profile}a), the domain wall position and azimuthal angle can be extracted from the simulation data, which match well with the analytical results [Eqs. (\ref{eq_profile}b) and (\ref{eq_profile}c)], as shown in Fig. \ref{fig1}(b). This justifies the omission of local demagnetization energy around the domain wall in Eq. (\ref{eq_energy}).

We first investigate the DWSK motion driven by the STT, which is generated by injecting an in-plane electric current into the ferromagnetic strip and has the form
\begin{equation}\label{eq_STT}
  \tau_{\mathrm{STT}}=-(\mathbf{u}\cdot\nabla)\mathbf{m}+\beta[\mathbf{m}\times(\mathbf{u}\cdot\nabla)\mathbf{m}],
\end{equation}
where $\mathbf{u}=-\mu_{B}P\mathbf{j}/[eM_{s}(1+\beta^2)]$, in which $\mu_{B}$ is the Bohr magneton, $P$ is the spin polarization, $\mathbf{j}$ is the current density, $e$ is the electron charge, and $\beta$ is the nonadiabatic coefficient.
Considering a rigid DWSK, its motion equation can be obtained by using the Thiele's collective coordinate approach, which is given as
\begin{equation}\label{eq_STT_Thiele}
  \mathbf{G}\times(\mathbf{v}-\mathbf{u})-\mathcal{D}(\alpha\mathbf{v}-\beta\mathbf{u})+\mathbf{F}=0,
\end{equation}
where $\mathbf{G}=-G \hat{z}$ with $G=4\pi Q$ is the gyrovector, $\mathcal{D}$ is the dissipation tensor defined as $D_{ij}=\iint(\partial_i\mathbf{m}\cdot\partial_j\mathbf{m})dxdy$, $\alpha$ is the Gilbert damping constant, $\mathbf{v}$ is the velocity of the DWSK, and $\mathbf{F}=-\nabla U$ is the force originating from the domain wall restriction and boundary repulsion.
Based on Eq. (\ref{eq_STT_Thiele}), the DWSK velocity can be derived as
\begin{subequations}\label{eq_STT_vel}
\begin{align}
  v_x &= \frac{(\alpha\beta D_{xx}D_{yy}+G^2)u_x+(\beta-\alpha)GD_{yy}u_y-G\partial_y U}{\alpha^2D_{xx}D_{yy}+G^2}, \\
  v_y &= \frac{(\alpha-\beta)GD_{xx}u_x+(\alpha\beta D_{xx}D_{yy}+G^2)u_y-\alpha D_{xx}\partial_y U}{\alpha^2D_{xx}D_{yy}+G^2},
\end{align}\end{subequations}
where $D_{xx}=16\sqrt{\kappa}(1+\kappa/3)$ and $D_{yy}=2L_x/\lambda$, $u_x$ and $u_y$ correspond to the current injected along the $x$ and $y$ directions, respectively. Since the periodic boundary conditions are adopted in the $x$ direction, we only consider the potential along the $y$ direction ($\partial_y U$).

\begin{figure}
  \centering
  \includegraphics[width=0.5\textwidth]{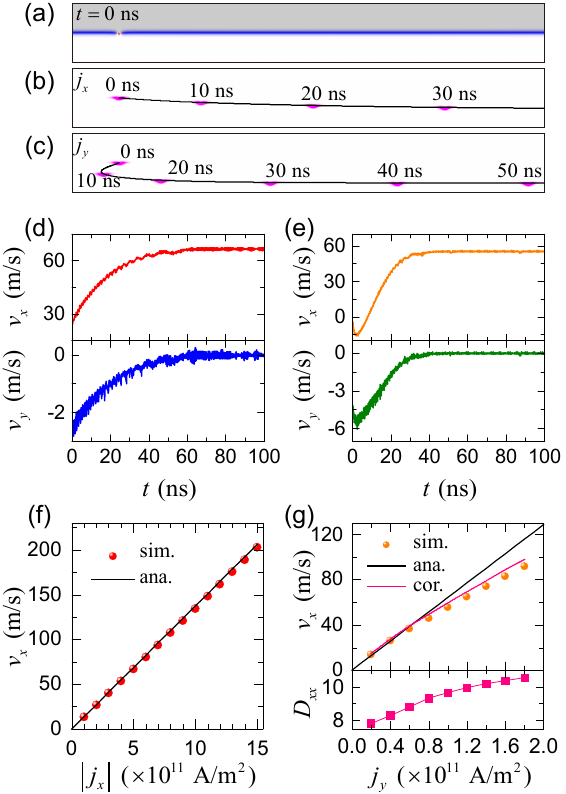}\\
  \caption{(a) The initial state of the DWSK with $Q=+1$. (b) and (c) The DWSK motion driven by the current injected along the $x$ and $y$ directions, respectively. The current densities in (b) and (c) are $j_x=-5\times10^{11}$ $\mathrm{A/m^{2}}$ and $j_y=1\times10^{11}$ $\mathrm{A/m^{2}}$, respectively. (d) and (e) The DWSK velocity as a function of the time in (b) and (c). (f) and (g) The dependence of the DWSK velocity on the current density for the current injected along the $x$ and $y$ directions, respectively. The bottom panel in (g) shows the dissipation tensor component $D_{xx}$ as a function of the current density obtained from simulations. In (f) and (g), the symbols are the simulation data and the lines are the analytical results. The pink line in (g) is the analytical curve with the correction of $D_{xx}$.}\label{fig2}
\end{figure}

In simulations, we set $P=0.5$ and $\beta=0.3$. The initial state is shown in Fig. \ref{fig2}(a), where the domain wall is positioned at the strip center and hosts a DWSK with $Q=+1$ on its left side. The position of the DWSK can be tracked by its guiding center, which is defined as
\begin{equation}\label{eq_pos}
  r_i=\frac{\int[i \mathbf{m}\cdot(\partial_x \mathbf{m}\times\partial_y \mathbf{m})]dxdy}{\int[\mathbf{m}\cdot(\partial_x \mathbf{m}\times\partial_y \mathbf{m})]dxdy}, \enspace i=x,y.
\end{equation}
%Then, the DWSK velocity can be calculated by $\mathbf{v}=(\dot{r}_x,\dot{r}_y)$.

Figure \ref{fig2}(b) illustrates the DWSK trajectory with a driving current in the $x$ direction (parallel to the domain wall) at $j_x=-5\times10^{11}$ $\mathrm{A/m^{2}}$. The DWSK moves longitudinally along the current direction and also in the $y$ direction due to the Magnus force from its topological charge ($Q=+1$).
This $y$-direction motion bends the domain wall and pulls it towards the boundary. In turn, the domain wall exerts the elastic force that hinders this motion, mitigating the skyrmion Hall effect.
As the DWSK approaches the boundary, the $y$-direction motion stops, and the transverse velocity drops to zero ($v_y=0$) due to the combined effects of the domain wall restriction and the boundary repulsion, as shown in Fig. \ref{fig2}(d). Ultimately, the DWSK moves at a constant velocity along the $x$ direction.

We also examine the motion of the DWSK when the current is applied perpendicular to the domain wall ($\hat{y}$), as shown in Fig. \ref{fig2}(c). Initially, the DWSK moves in the $-x$ direction. Subsequently, it reverses direction and moves towards the $+x$ direction.
From Eq. (\ref{eq_STT_vel}a), the $x$-direction velocity can be determined as $v_x=[(\beta-\alpha)GD_{yy}u_y-G\partial_yU]/(\alpha^2D_{xx}D_{yy}+G^2)$. Before 10 ns, the DWSK is distant from the boundary, the current's driving force prevails over the boundary's repulsive force (i.e. $|(\beta-\alpha)GD_{yy}u_y|>|G\partial_yU|$), resulting in $v_x<0$. As time passes, the DWSK gets closer to the boundary, and the boundary repulsion takes over ($|(\beta-\alpha)GD_{yy}u_y|<|G\partial_yU|$), causing $v_x$ to become positive.
Concurrently, the $y$-direction velocity decreases to zero as the DWSK approaches the boundary [see Fig. \ref{fig2}(e)]. Finally, the DWSK moves at a constant speed along the domain wall.
Additionally, the $y$-direction current can push the domain wall towards the boundary, making the DWSK more susceptible to annihilation.

Assuming the DWSK without the $y$-direction motion ($v_y=0$), Eq. (\ref{eq_STT_vel}) can be reduced to
\begin{equation}\label{eq_STTx_vx}
  v_x=\frac{\beta}{\alpha}u_x,
\end{equation}
for the current injected along the $x$ direction, and
\begin{equation}\label{eq_STTy_vx}
  v_x=-\frac{G}{\alpha D_{xx}}u_y,
\end{equation}
for the current injected along the $y$ direction. The steady velocities of the DWSK extracted from the simulations for different current densities are plotted in Figs. \ref{fig2}(f) and \ref{fig2}(g). Numerical results agree well with the analytical formula Eq. (\ref{eq_STTx_vx}) when the current is injected along the $x$ direction, but deviate for the $y$-direction current at $j_y\geq1\times10^{11}$ $\mathrm{A/m^{2}}$.
This deviation occurs because the domain wall bending becomes more pronounced under high current densities, resulting in an increase of the dissipation tensor component $D_{xx}$ [lower panel in Fig. \ref{fig2}(g)]. Consequently, the velocity obtained from simulations is lower than the analytical value. Adjusting $D_{xx}$ in Eq. (\ref{eq_STTy_vx}) by considering the domain wall bending can reduce this deviation [see the pink curve in the upper panel of Fig. \ref{fig2}(g)].
It is also found that the $x$-direction velocity ($v_x=-Gu_y/\alpha D_{xx}$) is proportional to the gyrovector $G$, indicating that the DWSK motion along the domain wall ($\hat{x}$) is induced by the skyrmion Hall effect.

Most recently, it was demonstrated that the DWSK can be driven by currents injected perpendicular to the domain wall \cite{Han2024}, where a perpendicular magnetic field is applied to eliminate the skyrmion Hall effect-induced motion along the domain wall. Consequently, the DWSK and domain wall move together along the current direction, which causes the domain wall to lose its functionality as a channel for guiding the DWSK motion. Instead, our work shows that the skyrmion Hall effect can be harnessed to drive the DWSK, an idea also evidenced in the movement of domain-wall bimeron driven by the SOT \cite{Chen2024}.

\begin{figure}
  \centering
  \includegraphics[width=0.5\textwidth]{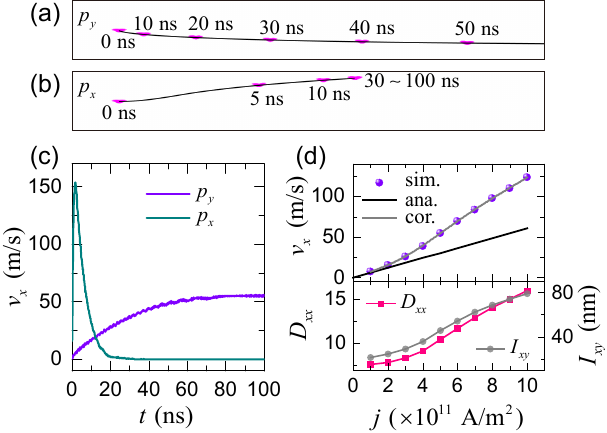}\\
  \caption{The DWSK motion driven by the SOT with the spin polarization (a) $p_y=-1$ and (b) $p_x=1$. The current density in both cases is $j=5\times10^{11}$ $\mathrm{A/m^{2}}$. In (b), the DWSK stops after 30 ns. (c) The correspond velocities of the DWSK in (a) and (b). (d) The dependence of the DWSK velocity (upper panel) and parameters of $D_{xx}$ and $I_{xy}$ (bottom panel) on the current density for the $y$-direction polarization. The symbols are the simulation data. The black and gray curves in the upper panel of (d) are the analytical results without and with the correction of $D_{xx}$ and $I_{xy}$.}\label{fig3}
\end{figure}

Next, we investigate the DWSK motion driven by the SOT, which can be generated by the spin-Hall-effect-induced spin currents from the adjacent heavy metal layer \cite{Liu2011}. Here, we only consider the damping-like term,
\begin{equation}\label{eq_SOT}
  \tau_{\mathrm{SOT}}=-a_j\mathbf{m}\times(\mathbf{m\times\mathbf{p}}),
\end{equation}
where $a_j=\gamma\hbar\theta_{\mathrm{SH}}j/(2eL_zM_s)$ with the gyromagnetic ratio $\gamma$, the reduced Planck constant $\hbar$, and the spin Hall angle $\theta_{\mathrm{SH}}$. The spin polarization vector $\mathbf{p}$ is oriented along the $y$ and $x$ direction for the current flowing parallel and perpendicular to the domain wall through the heavy metal layer, respectively. The Thiele equation, which describes the DWSK motion driven by the SOT, is written as
\begin{equation}\label{eq_SOT_Thiele}
  \mathbf{G}\times\mathbf{v}-\alpha \mathcal{D}\mathbf{v}-a_{j}\mathcal{I}\mathbf{p}=\nabla U,
\end{equation}
where $\mathcal{I}$ is the tensor defined as $I_{ij}=\iint(\partial_i \mathbf{m}\times\mathbf{m})_{j}dxdy$. From Eq. (\ref{eq_SOT_Thiele}), we can obtain the DWSK velocity
\begin{subequations}
  \begin{align}
    v_x &= \frac{-Ga_jI_{yx}p_x-\alpha D_{yy}a_jI_{xy}p_y-G\partial_yU}{\alpha^2D_{xx}D_{yy}+G^2}, \\
    v_y &= \frac{-\alpha D_{xx}a_jI_{yx}p_x+Ga_jI_{xy}p_y-\alpha D_{xx}\partial_yU}{\alpha^2D_{xx}D_{yy}+G^2},
  \end{align}\label{eq_SOT_vel}
\end{subequations}
where $I_{xy}=8\pi\sqrt{\kappa}\lambda/3$, $I_{yx}=-\pi(L_x+4\lambda/\sqrt{\kappa})$, and $p_y=-1$ and $p_x=1$ correspond to the current flowing through the heavy metal layer in the $x$ and $y$ directions, respectively.
Due to the boundary repulsion, we assume $v_y=0$ and obtain the steady velocity of the DWSK
\begin{equation}\label{eq_SOTy_vx}
  v_x=-\frac{a_jI_{xy}p_y}{\alpha D_{xx}},
\end{equation}
for the $y$-direction polarization, and $v_x=0$ for the $x$-direction polarization.

Figures \ref{fig3}(a) and \ref{fig3}(b) show the trajectory of the DWSK driven by the SOT obtained from the simulations, where we set $j=5\times10^{11}$ $\mathrm{A/m^{2}}$, and $\theta_{\mathrm{SH}}=0.05$. For $p_y=-1$, the DWSK initially moves towards the boundary and then begins to move along it with a constant velocity [see Figs. \ref{fig3}(a) and \ref{fig3}(c)], similar to the STT scenario with the $x$-direction current. Figure \ref{fig3}(d) presents the dependence of the DWSK velocity on the current density, revealing a significant discrepancy between numerical and analytical results. This is attributed to the domain wall deformation during the DWSK motion, leading to a notable change in $D_{xx}$ and $I_{xy}$ [shown in the lower panel of Fig. \ref{fig3}(d)]. Correcting $D_{xx}$ and $I_{xy}$ in Eq. (\ref{eq_SOTy_vx}) leads to better consistency between simulations and analytical results [see the purple dots and gray curve in the upper panel of Fig. \ref{fig3}(d)].
When the spin current is polarized in the $x$ direction ($p_x$), the DWSK motion differs from the case of the $p_y$ polarization, as shown in Figs. \ref{fig3}(b) and \ref{fig3}(c). It can be seen that the DWSK moves rapidly towards the boundary and then stops, confirming our theoretical prediction.

\begin{figure*}[htb]
  \centering
  \includegraphics[width=\textwidth]{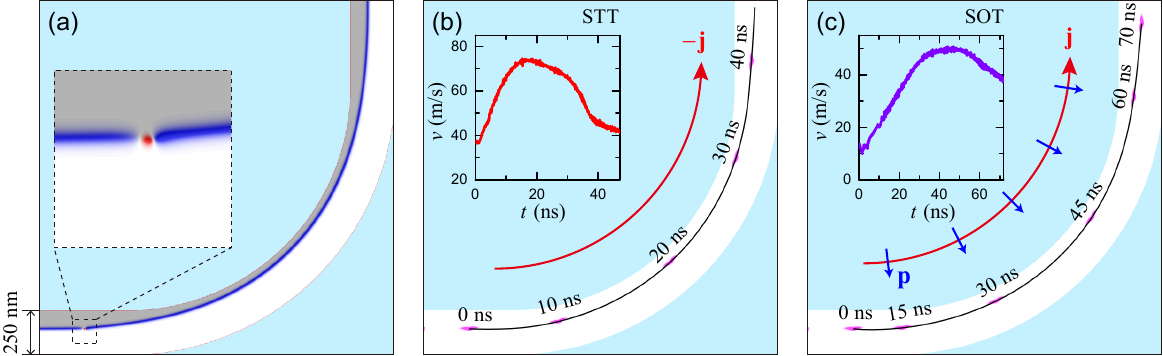}\\
  \caption{(a) The initial state of the DWSK in the curved domain wall. The inset shows the enlarge image of the DWSK. (b) and (c) The DWSK motion along the curved domain wall driven by the STT and SOT, respectively. The current density in both cases is $j=5\times10^{11}$ $\mathrm{A/m^{2}}$. The red curved arrows show the current flowing direction. The blue arrows in (c) are the spin polarization directions. The insets in (b) and (c) are the instantaneous velocities of the DWSK.}\label{fig4}
\end{figure*}

The current-driven motion of the DWSK in the curved domain wall is explored by micromagnetic simulations, as shown in Fig. \ref{fig4}. A 250 nm wide curved strip featuring a $90^{\circ}$ bend is considered, with an outer radius of 1500 nm. Figure \ref{fig4}(a) shows the initial state, which contains a domain wall with a DWSK located at its bottom left corner. The current is injected along the curved strip, parallel to the curved domain wall, as indicated by red arrows in Figs. \ref{fig4}(b) and \ref{fig4}(c). The polarization direction of the SOT is established by $\mathbf{p}=\mathbf{j}\times \hat{z}$, as illustrated by blue arrows in Fig. \ref{fig4}(c). It is evident that the DWSK can moves smoothly along the curved strip, driven by the current through either STT or SOT. The speed of the DWSK in the curved domain wall is also indicated in the insets of Figs. \ref{fig4}(b) and \ref{fig4}(c).
%Furthermore, it is noted that the speed of the DWSK moving in the curved portion is higher than that in the straight portion, as evidenced by the insets of Figs. \ref{fig4}(b) and \ref{fig4}(c). The velocity of the DWSK motion in the straight strip is also provided for comparison.

\begin{figure}[htb]
  \centering
  \includegraphics[width=0.5\textwidth]{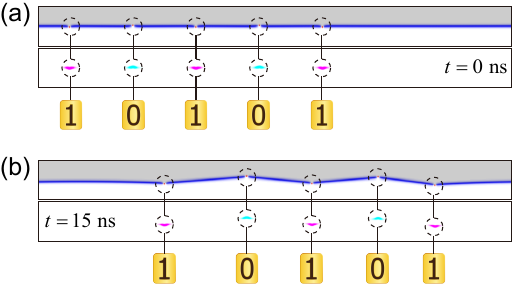}\\
  \caption{(a) The representation of data information in the DWSK-based racetrack, where the data bits ``1" and ``0" are encoded by the DWSKs with $Q=\pm1$, respectively. (b) The transmission of data information (chain of five DWSKs) driven by the STT with $j_x=5\times10^{11}$ $\mathrm{A/m^2}$. In (a) and (b), the upper and bottom panels are the magnetization state and topological charge density of the DWSK chain.}\label{fig5}
\end{figure}

The above results suggest that domain walls can support the DWSKs with opposite topological charges and act as a channel for guiding their motion. Therefore, the DWSKs are suitable for usage as information carriers in racetrack memory, where data bits ``1" and ``0" are encoded by DWSKs with $Q=\pm1$, as depicted in Fig. \ref{fig5}(a). Information transmission can be achieved by the motion of DWSKs driven by the spin-polarized current. Figure \ref{fig5}(b) presents a snapshot of the DWSK motion driven by the STT with the $x$-direction current. One can see that the spacing between the DWSKs varies, potentially due to the interaction between DWSKs and the domain wall deformation. However, the information fidelity is preserved because data bits are encoded by the DWSKs with opposite topological charges, rather than by the presence or absence of skyrmions as in the skyrmion-based racetrack memory. Moreover, the DWSKs with $Q=\pm1$ move in opposite directions, downwards and upwards respectively, providing a identifier for information reading. Our design principle is also applicable to other domain-wall solitons. For example, Bloch lines \cite{Malozemoff1972} in domain walls can be utilized to design a type of racetrack similar to domain wall-based racetrack memories \cite{Parkin2008}.

%Practical devices based on the DWSKs require the creation, manipulation, and detection of the DWSKs. It has been numerically demonstrated that the DWSK can be created by magnetic fields \cite{Kim2023} or by capturing a magnetic skyrmion into the domain wall \cite{Gudnason2024}.
%The DWSK can be driven by the spin-polarized current and magnetic field \cite{Han2024}. Although the skyrmion Hall effect is partially suppressed by the domain wall, under a large current density, the DWSK can bend the domain wall and drag it towards the boundary, leading to the annihilation of the DWSK. Fortunately, the shape and position of the domain wall can be fixed by a spatially-engineered DMI \cite{Mulkers2017} or magnetocrystalline anisotropy \cite{Yu2023}, which can efficiently prevent the DWSK annihilation.
%Future investigations on the generation and manipulation of the DWSK can utilize other methods, including spin waves \cite{Zhang2017,Wang202002}, surface acoustic waves \cite{Rivelles2024,Yang2024}, and thermal gradients \cite{Kong2013,Wang2020}, as well.
%In experiments, the DWSKs can be imaged using the Lorentz transmission electron microscopy \cite{Cheng2019,Li2021,Nagase2021} or nitrogen-vacancy center magnetometry \cite{Tetienne2014}.

In conclusion, our investigation into the current-driven motion of DWSKs within the ferromagnetic nanostrip reveals that for the STT, DWSKs can move steadily under currents injected either parallel or perpendicular to the domain wall. However, the motion of DWSKs is only achievable through the SOT when the polarization is in the $y$-direction, and it ceases when the polarization is in the $x$-direction. Apart from straight nanostrips, DWSKs can also traverse along curved domain walls. Interestingly, DWSKs with opposite topological charges can coexist within a single domain wall, which serves as a track to guide their motion. Based on these findings, we propose a DWSK-based racetrack memory, where information can be transmitted through the current-driven motion of DWSKs. Our results contribute to the understanding of DWSK motion induced by spin-polarized currents and could advance racetrack memory technologies.
\\

We thank Z. Zeng for helpful discussions. This work is supported by the Fundamental Research Funds for the Central Universities. Z.W. acknowledges the support the Natural Science Foundation of China (NSFC) (Grant No. 12204089) and the Natural Science Foundation of Hunan Province of China (Grant No. 2024JJ6113).
Z.-X.L. acknowledges financial support from the NSFC (Grant No. 11904048) and the Natural Science Foundation of Hunan Province of China (Grant No. 2023JJ40694).
X.S.W. acknowledges the support from the NSFC (Grant No. 12174093).

%%参考文献

\end{document}